\documentclass{article}
\usepackage{spconf,amsmath,graphicx}

\usepackage{hyperref}
\usepackage{todonotes}
\usepackage{subcaption}
\usepackage[normalem]{ulem} 

\title{Identifying differences in physical activity and autonomic function patterns between psychotic patients and controls over a long period of continuous monitoring using wearable sensors \vspace{-0.2cm}}
\usepackage{booktabs}

\name{
  \begin{tabular}[t]{@{}l@{}}
P. P. Filntisis$^1$, A. Zlatintsi$^1$, N. Efthymiou$^1$, E. Kalisperakis$^{2,3}$, T. Karantinos$^{2}$, M. Lazaridi$^{2,3}$, \\ 
\textit{\hspace{5cm} N. Smyrnis$^{2,3}$ and P. Maragos$^1$}\thanks{This research has been financed by the European Regional Development Fund of the European Union and Greek national funds through the Operational Program Competitiveness, Entrepreneurship and Innovation, under the call RESEARCH–CREATE–INNOVATE (project code:T1EDK-02890).}
\end{tabular} \vspace{-0.5cm}} 

\address{\selectfont\small$^1$School of ECE, National Technical University of Athens, 15773 Athens, Greece\\
\selectfont\small$^2$Laboratory of Cognitive Neuroscience, University Mental Health Research Institute, Athens, Greece\\
\selectfont\small$^3$National \& Kapodistrian University of Athens, Medical School\\
\vspace{1pt}
{\fontfamily{tt}\selectfont\small
\{filby,nefthymiou\}@central.ntua.gr, \{nzlat, maragos\}@cs.ntua.gr, smyrnis@med.uoa.gr
}
}

\vspace{-0.15cm}
\begin{document}
\ninept
\maketitle
\begin{abstract}
Digital phenotyping is a nascent multidisciplinary field that has the potential to revolutionize psychiatry and its clinical practice.
In this paper, we present a rigorous statistical analysis of short-time features extracted from wearable data, during long-term continuous monitoring of patients with psychotic disorders and healthy control counterparts. Our novel analysis identifies features that fluctuate significantly between the two groups, and offers insights on several factors that differentiate them, which could be leveraged in the future for relapse prevention and individualized assistance.
\end{abstract}
\vspace{-0.2cm}
\section{Introduction}
\label{sec:intro}
\vspace{-0.2cm}
Wearable consumer products, such as smartwatches and fitness trackers, are gaining popularity every day and the enormous technological advances made in recent years have enabled reliable, unobtrusive and remote personalized collection of numerous behavioral and biometric signals through their sensors~\cite{patel2012review, boletsis2015use}. 

This so called ``digital phenotyping"~\cite{torous} has enabled significant advances in wearables for health purposes, leading to the fact that next-generation wearable technologies are about to help transform nowadays hospital-centered healthcare practice to proactive, individualized care. Behavioral and biometric indexes have been already used in general medicine and sports and nowadays the evidence indicates that they could be introduced into clinical psychiatry \cite{aung2017sensing}, as well. Despite extensive research over the last 60 years in neurobiology and neurophysiology of psychotic disorders, their cause remains unclear and reliable biometric indexes for the diagnosis and prediction of the course of the psychotic symptomatology have not yet been found. The use of such signals for the detection of early diagnosis and prevention of psychotic relapses is now one of the major research areas in psychiatry \cite{EBM95, koutsouleris2011early, mcgorry2014biomarkers}. 




The e-Prevention project\footnote{More info can be found at: \href{http://eprevention.gr}{http://eprevention.gr}} is an ongoing research and development project with the goal of collecting long-term continuous recordings of biometric and behavioral signals through non-intrusive commercial wearable sensors (i.e., smartwatches), in order to develop innovative, advanced and valuable tools. Such tools would facilitate the effective monitoring, the prediction of clinical symptoms and the identification of biomarkers, which correlate with behavioral changes in patients with psychosis so as to support the relapse prevention. Timely detection of such relapses is in fact of major importance, not only for the clinicians; since patients not often present themselves
when the symptoms begin to re-emerge or worsen \cite{ corrigan1990noncompliance}, but it could also assist in reducing the severity of the relapses or even prevent their occurrences.

In contrast with previous works, which have lasted from some hours to a few weeks~\cite{VNL14, BTS18, cella2018using}, with some exceptions~(\cite{adler2020predicting}), our ongoing research study has been going on for more than one year, with the goal of achieving two years of continuous monitoring. In addition, previous works have mostly used smartphones~\cite{reyes2014human}, and  focused mainly in social features such as text messages, call duration or other such as location data, screen on/off time, and sleep duration.~\cite{BTS18, ben2017crosscheck, adler2020predicting}. Compared to smartphones, wearable sensors 
are unobtrusive, lightweight and can be used 
for monitoring while the subjects perform daily activities~\cite{ mukhopadhyay2014wearable}, ensuring this way a safe and sound living environment. Additionally, it has been already shown that people with psychotic illnesses are comfortable, able and willing to use personal digital devices to monitor outcomes in their daily life, supporting the fact that by using wearable sensors we could go beyond feasibility and underscore the novel physiological and activity data that can be easily collected with low cost~\cite{ robotham2016we, staplescomparison}. 

In our work we employ a commercial off-the-shelf smartwatch, aimed to have minuscule intrusion in the subject's life and be worn 24/7 (except during charging). The nature of our long-term study asks for a different data processing approach than previous studies. Inspired by traditional signal processing techniques, we extract common and more complex features using short-time analysis, and study them 
through their descriptive statistics in order to obtain a rough estimate of how they differentiate between healthy controls and patients with psychotic disorders. {\color{black} The experimental evaluation shows that both the more common, but also some of the novel nonlinear features examined are powerful in discriminating between the two groups.} The analysis conducted in this work is a vital step towards developing a method that can leverage physiological and behavioral data from sensors in order to timely predict relapses or adverse drug reactions.

\vspace{-0.2cm}
\section{Experimental Protocol and Data Collection}
\label{sec:eprev-db}
\vspace{-0.1cm}

\subsection{Experimental Protocol}
Twenty-three (23) healthy control volunteers and 22 patients with a disorder in the psychotic spectrum (9 with Schizophrenia, 8 with Bipolar Disorder I, 3 with Brief Psychotic Episode and 2 with Schizoaffective disorder) were recruited at the University Mental Health, Neurosciences and Precision Medicine Research Institute ``Costas Stefanis" (UMHRI) in Athens, Greece. All volunteers gave written consent for their participation after being fully informed about the project and also written permission for the use of their personal data (anonymized), in accordance with the provisions of the General Regulation (EU) 2016/679. Additionally, all protocols of the research project have been approved by the Ethics Committee of the Institution. 

Initially, the controls 
underwent a clinical evaluation to ensure there was no history of mental disorders or toxic substance usage, while for the recruitment of the patients, the clinicians met with the participants to conduct assessment of symptoms and functioning. At recruitment, patients were in active treatment and stable. The clinical team also conducted follow-up assessments with patients once every month of the study to administer various reliable rating scales (i.e., PANSS - Positive and Negative Syndrome Scale), which measure various psychiatric symptoms associated with their psychosis. 

Table~\ref{tab:demographics} contains information on the demographics of the two groups as well as the collected data (
described in Sec.~\ref{sec:3}) at the time of writing this paper. We also include the BMI (Body-Mass index) and the PANSS scale rating at the time of recruitment for the two groups (PANSS only applicable to patients).

\begin{table}[t]
\footnotesize
\centering
\begin{tabular}{l|c|c}
                           & Controls    & Patients     \\ \hline\hline
\textbf{Demographics}  & & \\ 
Male/Female                & 12/11       & 15/7         \\
Age (years)                & 27.8 $\pm$ 3.9   & 31.0 $\pm$ 6.21    \\
Education (years)          & 16.9 $\pm$ 1.8   & 14.04 $\pm$ 2.3    \\
Smoker/Non-smoker         & 4/19        & 15/7\\
Illness dur. (years)   & -           & 7.45  $\pm$  5.89   \\
BMI                        & 22.9  $\pm$  3.2 & 28.23  $\pm$  5.31  \\
PANSS (overall)                      & -           & 57.7 $\pm$ 13.9   \\ \hline\hline
\textbf{Recorded Data}  & & \\
\# Weeks Recorded          & 20.17 $\pm$ 5.27 & 17.4 $\pm$ 6.6   \\
\# 10 min. mov (awake)                        & 7780 $\pm$ 2385 & 5963 $\pm$ 2646 \\
\# 1 hour HRV (awake)                        & 858 $\pm$ 234 & 743 $\pm$ 367 \\
\# 10 min. mov (sleep)                        & 3767 $\pm$ 1282 & 4058 $\pm$ 1958 \\
\# 1 hour HRV (sleep)                        & 460 $\pm$ 164 & 551 $\pm$ 290 \\
\end{tabular}
\caption{Demographics information of controls and patients at the time of recruitment, illness information, and amount of recorded data for each group during wakefulness and sleep. There were no significant differences for the recorded data.}
\label{tab:demographics}
\vspace{-0.5cm}
\end{table}

\vspace{-0.25cm}
\subsection{Method \& Data Collection}
The subjects wore a Samsung Gear S3 smartwatch that continuously monitored acceleration (\textit{acc}), angular velocity (\textit{gyr}), and the heart rate (via Photoplethysmography~\cite{allen2007photoplethysmography}). Due to limits on the number of available devices, each subject was recruited at a different date - controls were recruited between June 2019 and October 2019 while patients have been continuously recruited from November 2019 up to now (Oct.~2020). Controls were continuously monitored for at least $90$ days and then returned the watches, while the monitoring of patients is an ongoing process. 
In the analysis presented in this paper, to mitigate the effect of the CoVID\-19 Pandemic quarantine lockdown (15/03--10/05/2020 in Greece), we exclude data collected during this period.

\begin{figure}[!th]
\centering
\includegraphics[width=0.5\textwidth]{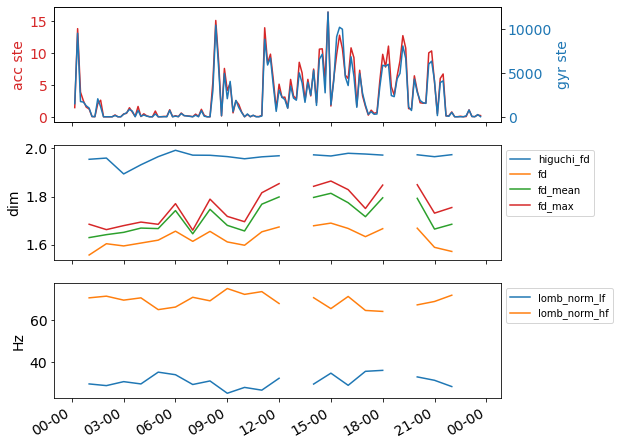}
\caption{Example of features considered in this paper during one subject's day. Top shows \textit{acc} and \textit{gyr} STE, middle shows fractal dimensions of HRV (Higuchi and MFD), and bottom figure shows HRV normalized LF and HF power extracted with LS. Missing values denote periods where HRV 
could not be detected correctly.}
\label{fig:signals}
 \vspace{-0.3cm}
\end{figure}

Data were collected using an in-house developed application and uploaded every day to a secure cloud server~\cite{maglogiannis2020intelligent}. Accelerometer (\textit{acc}) and gyroscope data (\textit{gyr}) were collected at a frequency of 20Hz, while the heart rate and the heart rate variability (RR intervals -- time intervals between two successive heart pulses) were collected at a rate of 5Hz (if a new beat was not detected the watch duplicates the last obtained value). Using the Tizen API provided by the smartwatch, we also collected information about the sleep schedule of the subjects, and their steps at aggregated intervals of 10 minutes.

\begin{figure*}[h]
\centering

\begin{subfigure}{1\textwidth}
    \centering
    \includegraphics[width=\textwidth]{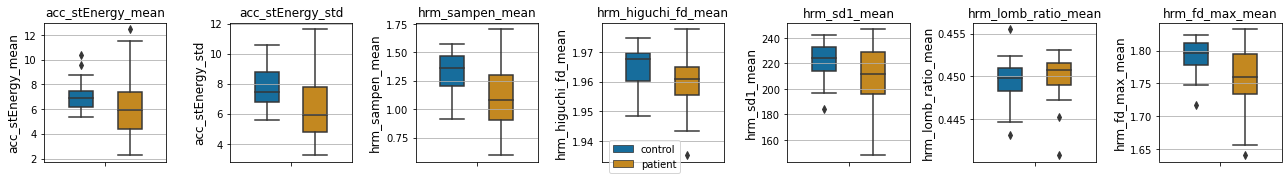}
\end{subfigure}%

\begin{subfigure}{1\textwidth}
    \centering
    \includegraphics[width=\textwidth]{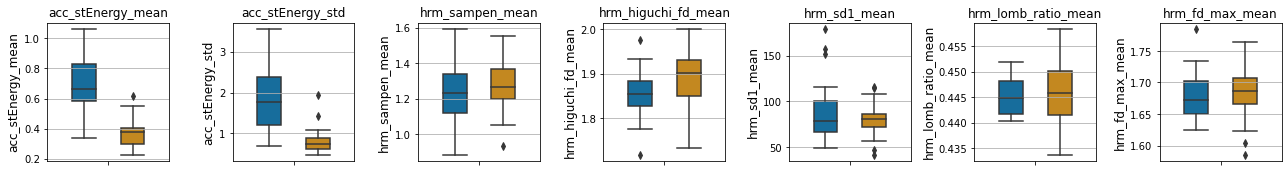}
\end{subfigure}
\caption{Boxplots for features of controls and patients while awake (top row) and asleep (bottom-row). The bold line represents the median, the boxes extend between the 1st and 3rd quartile, whiskers extend to the lowest and highest datum within 1.5 times the inter-quantile range (IQR) of the 1st and 3rd quartile respectively, and outliers are shown as diamonds.}
\label{fig:violin}
\vspace{-0.3cm}
\end{figure*}

\vspace{-0.20cm}
\section{Data Processing and Feature Extraction}
\label{sec:3}
\subsection{Data Preprocessing}
The heart rate variability (HRV) sequence from the 5Hz signal was obtained by dropping identical consecutive values and ensuring that for each 1 hour interval, the obtained sequence of RR intervals was summing up to at least 54 minutes (an empirical threshold corresponding to 90\% of valid heart data). We also removed RR intervals larger than 2000ms and smaller than 300ms as artifacts and replaced possible non-detected pulses with linear interpolation. We did not perform any kind of noise reduction in \textit{acc} and \textit{gyr}, since we determined that for the examined features the effect of noise is negligible.

\vspace{-0.25cm}
\subsection{Feature Extraction}
Short-time analysis of signals using windowing is a traditional signal processing method. In short-time analysis we assume the process under which the data are generated to be stationary. Drawing power from these techniques, but largely increasing the time scale, we proceeded to perform ``short-time" analysis in windows of 10 minutes for movement data (\textit{acc}, \textit{gyr}) and 1 hour for HRV. {\color{black}The 10 minutes intervals that were chosen for the analysis of the movement data and the aggregation of the features have been found optimal for distinguishing short-term patterns in a previous study~\cite{retsinas2020person}.} The mean and standard deviation of the number of intervals for each user is reported in Table~\ref{tab:demographics}. We consider the following features:

\textbf{Energy}
The energy (STE) of the euclidean norm of \textit{acc} and \textit{gyr} is extracted (since they are measured triaxially). We use these features as an objective measure of physical activity and general movement behavior. 


\textbf{Spectral features} Medical studies split the HRV spectrum in four frequency bands: ultra-low-frequency (ULF $\leq 0.003$ Hz), very-low-frequency (VLF $0.0033$–$0.04$ Hz), low-frequency (LF $0.04$–$0.15$ Hz), and high-frequency (HF $0.15$–$0.40$ Hz)\cite{shaffer2017overview}. Since HRV is by definition a non-uniformly sampled signal we perform spectral analysis using the Lomb-Scargle periodogram~\cite{scargle}, and we extract for each interval the relative power and normalized power in two bands: LF and HF, as well as the ratio LF-to-HF.

\textbf{Sample Entropy}
Nonlinear methods treat the extracted time series as the output of a nonlinear system. A typical characteristic of a nonlinear system is its complexity. The first measure of complexity we consider is the sample entropy (SampEn). Sample entropy is a measure of the rate of information generation by the system, considered an improvement over approximate entropy \cite{richman2000physiological} due to its unbiased nature.

\textbf{Higuchi Fractal Dimension}
Multiple algorithms have been proposed for measuring the fractal dimensions of a time series. Here we use the Higuchi fractal dimension~\cite{higuchi1988approach}, which has been used extensively in neurophysiology due to its simplicity and speed.

{\color{black}
\textbf{Multiscale Fractal Dimension}
(MFD) is an efficient algorithm \cite{Mara94} that measures the short-time fractal dimension, based on the Minkowski-Bouligand dimension~\cite{Falk03}. Real-world signals do not have the same structure over different time scales; and by measuring the MFD we are able to examine the complexity and fragmentation of the signals at multiple scales, thus creating a profile of local MFDs at each time location. For this reason, we summarized the short-time measured MFD profiles by taking the following statistics: fd[1] (the fractal dimension), min, max, mean, and std for the 1 hour HRV data.} 

\textbf{Poincare plot measures}
The Poincare plot~\cite{brennan2001existing} is a kind of recurrence plot where each sample of a time series is plotted against the previous, and then an ellipse is fitted on this scatter plot. The width of the ellipse (SD1) is a measure of short-term HRV, while the length (SD2) is a measure of long-term HRV.

\textbf{Feature Aggregation}
Using the information on the sleep schedule of each subject we split the intervals in two groups -- one corresponding to intervals during sleep and one during wakefulness. We then calculated 
the mean and standard deviation (std) over all its intervals, resulting 
in 2 values for each subject and feature type; resulting in a total of 28 features.

\textbf{Sleep/Wake Ratio and Steps}
In addition to the above features, we also 
extracted for each subject the mean and standard deviation of his sleep/wake ratio, and mean number of steps each day. Since the number of recorded hours each day fluctuates, for these features only we keep data from subjects 
that have at least 30 days with 20 hours or more recorded (21 controls with 65 $\pm$ 26 days and 14 patients with 56 $\pm$ 19 days with no significant difference($p=0.18$).

Fig.~\ref{fig:signals} shows examples of STE, and HRV fractal dimensions and LF/HF frequencies, during one day of monitoring a subject.


\vspace{-0.15cm}
\section{Experimental Results}
\label{sec:experimental_results}

\begin{table*}[t]
\centering
\footnotesize
\begin{tabular}{l|llc|llc} 
                                  & \multicolumn{6}{c}{State}                                                                                                                                    \\ 
\hline 
                                  & \multicolumn{3}{c|}{Awake}                                                                                                                                  & \multicolumn{3}{c}{Sleeping}                                                                                                                              \\ 
\hline 
feature              & \multicolumn{1}{c|}{Controls}  & \multicolumn{1}{c|}{Patients}       & p value         & \multicolumn{1}{c|}{Controls}      & \multicolumn{1}{c|}{Patients}     & p value                 \\ 
\hline \hline
acc STE mean                      & 6.87 (1.24)                                             & \begin{tabular}[c]{@{}l@{}}5.944 (2.983)\\\end{tabular}       & 0.08                              & 0.666 (0.247)                                           & \begin{tabular}[c]{@{}l@{}}0.375 (0.109)\\\end{tabular}     & $\boldsymbol{<0.001}$              \\
acc STE std                       & \textbf{7.48 (2.05)}                                    & \textbf{5.911 (2.979)}                                        & \textbf{0.02}                     & 1.770 (1.168)                                           & \begin{tabular}[c]{@{}l@{}}0.750 (0.248)\\\end{tabular}     & $\boldsymbol{<0.001}$              \\
gyr STE mean                      & 4166 (1190)                                       & \begin{tabular}[c]{@{}l@{}}3603 (2269)\\\end{tabular} & 0.06                              & 344 (195)                                       & \begin{tabular}[c]{@{}l@{}}157 (95)\\\end{tabular}  & $\boldsymbol{<0.001}$              \\
gyr STE std                       & \textbf{5213 (2225)}                              & \textbf{3686 (2306)}                                  & \textbf{0.02 }                    & 1058 (876)                                      & \begin{tabular}[c]{@{}l@{}}376 (140)\\\end{tabular} & $\boldsymbol{<0.001}$              \\
HRV SampEn mean                   & \textbf{1.36 (0.26)}                                    & \textbf{1.081 (0.389)}                                        & \textbf{0.02}                     & 1.233 (0.217)                                           & \begin{tabular}[c]{@{}l@{}}1.267 (0.172)\\\end{tabular}     & 0.26                               \\
HRV SampEn std                    & 0.33 (0.10)                                             & \begin{tabular}[c]{@{}l@{}}0.370 (0.066)\\\end{tabular}       & 0.09                              & \textbf{0.218 (0.043)  }                                         & \begin{tabular}[c]{@{}l@{}}\textbf{0.271 (0.075)}\\\end{tabular}     & $\boldsymbol{<0.001}$              \\
HRV Higuchi mean                  & 1.97 (0.01)                                             & \begin{tabular}[c]{@{}l@{}}1.961 (0.009)\\\end{tabular}       & 0.06                              & \textbf{1.854 (0.055)}                                  & \textbf{1.901 (0.081)}                                      & \textbf{0.04}                      \\
HRV Higuchi std                   & 0.016 (0.004)                                           & \begin{tabular}[c]{@{}l@{}}0.018 (0.010)\\\end{tabular}       & 0.28                              & 0.056 (0.019)                                           & \begin{tabular}[c]{@{}l@{}}0.048 (0.023)\\\end{tabular}     & 0.31                               \\
HRV SD1 mean                      & 224.54 (18.84)                                          & \begin{tabular}[c]{@{}l@{}}211.508 (33.165)\\\end{tabular}    & 0.09                              & 78.431 (34.118)                                         & \begin{tabular}[c]{@{}l@{}}80.482 (15.065)\\\end{tabular}   & 0.35                               \\
HRV SD1 std                       & \textbf{29.39 (8.58)}                                   & \textbf{33.758 (10.080)}                                      & \textbf{0.02 }                    & 39.104 (23.996)                                         & \begin{tabular}[c]{@{}l@{}}40.162 (12.584)\\\end{tabular}   & 0.35                               \\
HRV SD2 mean                      & 262.98 (26.38)                                          & \begin{tabular}[c]{@{}l@{}}251.593 (46.782)\\\end{tabular}    & 0.13                              & 130.441 (38.627)                                        & \begin{tabular}[c]{@{}l@{}}127.220 (35.875)\\\end{tabular}  & 0.25                               \\
HRV SD2 std                       & \textbf{33.34 (6.54)}                                   & \textbf{39.492 (10.849)}                                      & \textbf{0.02}                     & 43.139 (23.254)                                         & \begin{tabular}[c]{@{}l@{}}50.610 (19.142)\\\end{tabular}   & 0.19                               \\
HRV HF power mean                 & 69.11 (0.12)                                            & \begin{tabular}[c]{@{}l@{}}69.074 (0.122)\\\end{tabular}      & 0.17                              & 69.362 (0.297)                                          & \begin{tabular}[c]{@{}l@{}}69.311 (0.424)\\\end{tabular}    & 0.34                               \\
HRV HF power std                  & 3.125 (0.095)                                           & \begin{tabular}[c]{@{}l@{}}3.093 (0.135)\\\end{tabular}       & 0.16                              & 3.157 (0.132)                                           & \begin{tabular}[c]{@{}l@{}}3.128 (0.142)\\\end{tabular}     & 0.34                               \\ 
{HRV MFD mean}   & \begin{tabular}[c]{@{}l@{}}1.637 (0.032)\\\end{tabular} & \begin{tabular}[c]{@{}l@{}}1.622 (0.037)\\\end{tabular}       & 0.12         & \begin{tabular}[c]{@{}l@{}}1.566 (0.053)\\\end{tabular} & \begin{tabular}[c]{@{}l@{}}1.582 (0.073)\\\end{tabular}     & 0.24          \\
{HRV MFD std}    & \begin{tabular}[c]{@{}l@{}}0.039 (0.011)\\\end{tabular} & \begin{tabular}[c]{@{}l@{}}0.043 (0.011)\\\end{tabular}       & 0.14          & \textbf{0.039 (0.010)}                                  & \textbf{0.047 (0.006)}                                      & \textbf{0.03}  \\
{HRV MFD max mean}  & \textbf{1.796 (0.034)}                                  & \textbf{1.761 (0.060)}                                        & \textbf{0.05} & \begin{tabular}[c]{@{}l@{}}1.673 (0.052)\\\end{tabular} & 1.686 (0.041)                                               & 0.31           \\
{HRV MFD max std}   & \textbf{0.051 (0.016)}                                  & \textbf{0.063 (0.019)}                                        & \textbf{0.04} & \textbf{0.049 (0.014)}                                  & \textbf{0.055 (0.009)}                                      & \textbf{0.04}  \\
{HRV MFD min mean}  & \begin{tabular}[c]{@{}l@{}}1.637 (0.032)\\\end{tabular} & \begin{tabular}[c]{@{}l@{}}1.622 (0.037)\\\end{tabular}       & 0.12          & \begin{tabular}[c]{@{}l@{}}1.566 (0.052)\\\end{tabular} & \begin{tabular}[c]{@{}l@{}}1.581 (0.069)\\\end{tabular}     & 0.24          \\
{HRV MFD min std}   & \begin{tabular}[c]{@{}l@{}}0.039 (0.011)\\\end{tabular} & \begin{tabular}[c]{@{}l@{}}0.043 (0.011)\\\end{tabular}       & 0.14          & \textbf{0.039 (0.010)}                                  & \textbf{0.047 (0.006)}                                      & \textbf{0.03}  \\
{HRV MFD mean mean} & \begin{tabular}[c]{@{}l@{}}1.745 (0.035)\\\end{tabular} & \begin{tabular}[c]{@{}l@{}}1.713 (0.057)\\\end{tabular}       & 0.06          & \begin{tabular}[c]{@{}l@{}}1.633 (0.050)\\\end{tabular} & \begin{tabular}[c]{@{}l@{}}0.047 (0.006)\\\end{tabular}     & 0.31          \\
{HRV MFD mean std}  & \begin{tabular}[c]{@{}l@{}}0.050 (0.014)\\\end{tabular} & \begin{tabular}[c]{@{}l@{}}0.057 (0.015)\\\end{tabular}       & 0.06          & \textbf{0.044 (0.013)}                                  & \textbf{0.050 (0.010)}                                      & \textbf{0.04}  \\
{HRV MFD std mean}  & \textbf{0.045 (0.003)}                                  & \textbf{0.040 (0.008)}                                        & \textbf{0.03} & \begin{tabular}[c]{@{}l@{}}0.030 (0.008)\\\end{tabular} & \begin{tabular}[c]{@{}l@{}}0.026 (0.006)\\\end{tabular}     & 0.15           \\
{HRV MFD std std}   & \textbf{0.007 (0.002)}                                  & \textbf{0.008 (0.003)}                                        & \textbf{0.03} & 0.009 (0.002)                                           & 0.010 (0.002)                                               & 0.051          \\
\bottomrule
\end{tabular}
\caption{Statistical difference analysis using Mann-Whitney U-tests with BH correction in each state. Bold values denote significance at the 95\% confidence levels. For each group the median and the IQR (in parenthesis) is shown for each feature.}
\label{tab:results}
\vspace{-0.25cm}
\end{table*}





\vspace{-0.25cm}
\subsection{Wakefulness comparison}
In Fig.~\ref{fig:violin} (top row) we show the boxplots of the features during wakefulness that were deemed more fitting (due to space limitations) to display differences between the two groups. Due to the differences observed perceptually 
between the distributions in most features, we tested for significant difference between distributions (the null hypothesis being that the two distributions are the same) using two-tailed non-parametric Mann-Whitney U tests~\cite{mann1947test}. We adjusted for p-values using the Benjamini-Hochberg (BH) procedure~\cite{benjamini1995controlling}. Due to the nature of our explorative study, BH was preferred over more strict Family-Wise Error Rate methods~\cite{chen2017general}.

Table~\ref{tab:results} shows the results of Mann-Whitney U tests for all features, while the subjects are awake (due to limited space we omit the values of LF and HF normalized powers which showed no significant differences). During wakefulness, the features that pertain to movements appear to present more variability in the patient group when compared to controls. The same appears to be true for some nonlinear HRV features (SampEn mean, Higuchi, SD1 mean, SD2, MFD max).
The testing showed significant distribution differences in the standard deviation of \textit{acc} and \textit{gyr} short-time energy, the \textit{standard deviation} of SD1 and SD2, the sampen mean and MFD max and std. The other features failed to reject the null hypothesis.

\vspace{-0.35cm}
\subsection{Sleep comparison}

Similarly, Fig.~\ref{fig:violin} (bottom row) presents the feature distributions for each group during sleeping. It is evident that especially the movement-related features present a significant difference, which is also verified in the Mann-Whitney U test results in Table~\ref{tab:results}. A similar result is found for the standard deviation in sample entropy of HRV as well as the mean of the Higuchi fractal dimension. MFD features (MFD fd[1] std, max std, min std, and mean std) were also found to differ significantly.

\begin{figure}[t]
\centering

\begin{subfigure}{.5\textwidth}
    \centering
    \includegraphics[width=\textwidth]{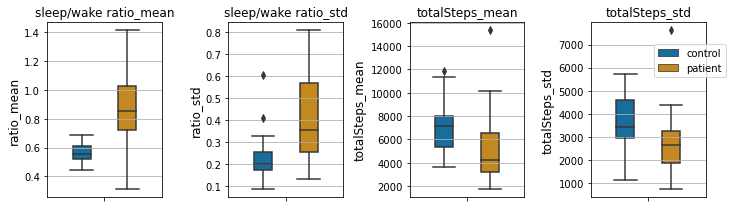}
\end{subfigure}%
\vspace{-0.25cm}
\caption{Boxplots of sleep/wake ratio and steps per day (mean-std).}
\label{fig:extra}
\vspace{-0.5cm}
\end{figure}

\vspace{-0.3cm}
\subsection{Sleep-wake ratio and total steps}
Finally, Fig.~\ref{fig:extra} shows the boxplots of the statistics of steps per day and sleep wake ratio for the two groups. We observe a large significant difference between both the distributions of the mean and std of the sleep wake ratio ($p<0.001$) as well as the distributions of total steps per day ($p<0.001$). 
\vspace{-0.35cm}

\section{Discussion}
\label{sec:discussion}
\vspace{-0.15cm}
Our goal with the statistical analysis in this work is to exploit traditional, but also less-known signal processing techniques to identify common markers/features that differ drastically when a person has a psychotic disorder. These markers could prove useful in predicting potential relapses in these patients. 

Our findings have shown that patients tend to behave with greater variability and present large outliers -- some behave close to controls, while others might show extreme values. 
During wakefulness, even though the mean energy did not differ when compared to controls, the standard deviation showed a significant difference, indicating that patients tend to depict large variations in their movement behavior. On the contrary, during sleeping the patients presented a small mean and standard deviation of the energy in each of their sleeping intervals compared to the controls. We should note however that the observed differences in sleep between the two groups could be attributed to medication administered to patients, which possibly causes variability in sleep duration as well.

{\color{black}Some of the nonlinear features that were measured for the HRV data} showed significant differences in the distributions between controls and patients, i.e., during sleeping, as seen in Table~\ref{tab:results}, such features are the standard deviation of the sample entropy, 
the mean of the Higuchi fractal dimension, {\color{black} as well as various statistics derived from the MFD analysis (i.e., min, max, and mean of the standard deviation measurement)}. During wakefulness, the mean of the sample entropy, the std of the poincare features sd1 and sd2 and again various statistics of the MFD presented significant results. Spectral analysis using the LS spectrogram did not show any significant differences between controls and patients.


The main merits of our work are two-fold: First, compared to previous similar studies, which have mostly lasted for a few weeks, our study has already been going on for more than a year with the goal to obtain two years of continuous monitoring of patients with psychotic disorders. To do this, we employ a commercial off-the-shelf smartwatch, that has been acknowledged by our volunteers to be comfortable and patients are willing to insert it into their daily lives routine. Second, we show how traditional short-time-analysis combined with common but also more complex {\color{black} and novel} features, {\color{black} such as the MFD features that depicted significant differences in both awake and sleeping data}, can be employed to identify biomarkers, present large inter-group variabilities between healthy controls and patients, paving a way towards both acquiring clinical insights on psychotic disorders, but also exploring the capabilities of these markers to predict relapses.

\vspace{-0.45cm}
\section{Conclusion}
\label{sec:conclusion}
\vspace{-0.15cm}
In this paper we identified markers that differentiate between healthy controls and people with psychotic disorders. To this end, we have specifically collected a large amount of physical activity and autonomic function data from wearable devices. 
Statistical analysis between the two groups, through their descriptive statistics, indicated significant differences regarding the movement behavior, 
as well as in some markers of cardiac function during both wakefulness and sleeping. In future analyses, we also intend to account for the effects of antipsychotics and/or other medications administered to patients, as well as other factors that differ in the two samples, such as smoker/non-smokers percentages. Finally, we aim to explore the capabilities of such markers to predict psychotic relapses and adverse drug effects. 

\bibliographystyle{IEEEbib}
\bibliography{refs}

\end{document}